\documentclass[two column,rsi,amsmath,amssymb,superscriptaddress,pre,floatfix] {revtex4-1}
\setcitestyle{super}
\usepackage{color,graphics}
\usepackage{graphicx}
\usepackage[sort&compress]{natbib}
\usepackage{hhline}
\usepackage{soul}
\usepackage{xcolor}
\usepackage{array}
\usepackage{bm}

\usepackage[normalem]{ulem}
\usepackage{hyperref}
\usepackage{natbib}

\usepackage[thinlines]{easytable}
\begin{document}
\setcounter{totalnumber}{3}
\renewcommand{\thetable}{\arabic{table}}
\newcolumntype{P}[1]{>{\centering\arraybackslash}p{#1}}

\title{Theoretical insights into the role of lattice fluctuations \\ on the excited behavior of lead halide perovskites}


\author{Yoonjae Park}
 \affiliation{Department of Chemistry, University of California, Berkeley, California 94720, USA}
 \affiliation{Materials Science Division, Lawrence Berkeley National Laboratory, Berkeley, California 94720, USA}

\author{Rohit Rana}
 \affiliation{Department of Chemistry, University of California, Berkeley, California 94720, USA}
 \affiliation{Materials Science Division, Lawrence Berkeley National Laboratory, Berkeley, California 94720, USA}

\author{Daniel Chabeda}
 \affiliation{Department of Chemistry, University of California, Berkeley, California 94720, USA}

\author{Eran Rabani}
 \affiliation{Department of Chemistry, University of California, Berkeley, California 94720, USA}
 \affiliation{Materials Science Division, Lawrence Berkeley National Laboratory, Berkeley, California 94720, USA}
 \affiliation{The Sackler Center for Computational Molecular and Materials Science, Tel Aviv University, Tel Aviv, 69978, Israel}

\author{David T. Limmer}
 \email{dlimmer@berkeley.edu}
 \affiliation{Department of Chemistry, University of California, Berkeley, California 94720, USA}
\affiliation{Materials Science Division, Lawrence Berkeley National Laboratory, Berkeley, California 94720, USA}
\affiliation{Chemical Science Division, Lawrence Berkeley National Laboratory, Berkeley, California 94720, USA}
\affiliation{Kavli Energy NanoScience Institute, Berkeley, California 94720, USA}

\date{\today}
\vspace{0mm}

\begin{abstract}
Lead halide perovskites have been extensively studied as a class of materials with unique optoelectronic properties. A fundamental aspect that governs optical and electronic behaviors within these materials is the intricate coupling between charges and their surrounding lattice. Unravelling the role of charge-lattice interactions on the optoelectronic properties in lead halide perovskites is necessary to understand their photophysics. Unlike traditional semiconductors where a harmonic approximation often suffices to capture lattice fluctuations, lead halide perovskites have a significant anharmonicity attributed from the rocking and tilting motions of inorganic framework. Thus, while there is broad consensus on the importance of the structural deformations and polar fluctuations on the behavior of charge carriers and quasiparticles, the strongly anharmonic nature of these fluctuations and their strong interactions render theoretical descriptions of lead halides perovskites challenging. In this Account, we review our recent efforts to understand how the soft, polar lattice of this class of materials alter their excited state properties. We  highlight the influence of the lattice on static properties by examining the quasiparticle binding energies and fine structure. With perovskite nanocrystals, we discuss how incorporating the lattice distortion is essential for accurately defining the exciton fine structure. By considering lattices across various dimensionalities, we are able to illustrate that the energetics of excitons and their complexes are significantly modulated by the polaron formation. Beyond energetics, we also delve into how the lattice impacts the dynamic properties of quasi-particles. The mobilities of charge carriers is  studied with various charge-lattice coupling models and the recombination rate calculation demonstrates the molecular origin on the peculiar feature in the lifetime of charge carriers in these materials. In addition, we address how lattice vibrations themselves relax upon excitation from charge-lattice coupling. Throughout, these examples are aimed at characterizing the interplay between lattice fluctuations and optoelectronic properties of lead halide perovskites and are reviewed in the context of the effective models we have built, and the novel theoretical methods we have developed to understand bulk crystalline materials, as well as nanostructures, and lower dimensionality lattices. By integrating theoretical advances with experimental observations, the perspective we detail in this account provides a comprehensive picture that serves as both design principles for optoelectronic materials and a set of theoretical tools to study them when charge-lattice interactions are important. 
\end{abstract}

\maketitle


\section{Introduction}

Over the past 20 years, lead halide perovskites have emerged as an important material for a variety of photovoltaic and light-emitting applications.
On top of their ease of synthesis, tunable optoelectronic properties and high defect tolerance, \cite{Zhu:2015eb4, sciadvzhu, protesescu2017} they have small exciton binding energies, \cite{herz2016charge} long carrier lifetimes and diffusion lengths \cite{science.aaa5333} despite modest charge mobilities, \cite{adma.201305172} and  have demonstrated high power conversion efficiencies and quantum yields. 
Consisting of a lead-halide inorganic framework and counter-balancing cations, lead halide perovskites are held together with ionic bonds that localize significant charge leading to strong Coulomb interactions between photoexcited carriers and the surrounding lattice. \cite{acs.jpclett.6b01425} The isotropic nature of this ionic bonding results in a deformable lattice, admitting structural fluctuations and polaron formation that can protect charges from scattering with defects, \cite{frohlich2, ptemass} and screen the interaction between charges. \cite{ncomms12253} 
Although many of the important optical properties in perovskites are presumed to arise from the interplay of electronic structure and charge-lattice interactions,\cite{sciadvzhu,limmer2020photoinduced}  elucidating their microscopic origin has remained challenging. 

The significant anharmonicity of perovskite lattices means that traditional models and perspectives on electron-phonon coupling are not  appropriate. Unlike most crystalline semiconducting materials where lattice fluctuations are well described by a harmonic approximation, perovskites have rocking and tilting motions of the inorganic framework that are better described by double well potentials, \cite{acs.jpclett.7b02423}  and they exhibit nearly free motions of their A-site cations \cite{physrevb.92.144308}. Unsurprisingly, these soft modes produce a number of competing structural phases that can be accessed over typical operating conditions or tuned through nanocrystalline size. In lower dimensional crystals, the A-site cation can be replaced with ionic ligands, which themselves can disorder and admit large fluctuations\cite{park2021}. Even at low temperatures, where a harmonic approximation may be valid, perovskites have complex unit cells, rendering the use of traditional theoretical methods computationally intractable. There have been a growing number of studies pointing to the importance of charge-lattice effects on the properties of lead-halide perovskites, motivating analytical \cite{multiphonon} and numerical studies \cite{multiphonon, mayers2018lattice}, but balancing an atomistic description of the lattice and a non-perturbative treatment of charge-lattice coupling has required that novel theoretical techniques be developed. We have developed path integral based methods in order to describe charge-lattice affects non-perturbatively \cite{park2022, parkjcp2022}, and adapted pseudo-potential methods \cite{10.1063/1.478431} to study nanocrystals and the role of surface relaxation on quasiparticle energetics and lifetimes \cite{10.1021/acsenergylett.9b02395}. These tools have enabled studies on exciton binding and charge carrier renormalization from polaron formation, and an elucidation of the interplay between lattice structure and exciton energies in nanomaterials.

In this Account, we summarize these theoretical developments and how they have been used to rationalize spectroscopic and excited state dynamics in these material systems.  In Sec.~\ref{static}, we present a discussion on static properties of excited states focusing on exciton fine structure and binding energies of excitons and biexcitons and how each depends on lattice geometry and particle morphology. We continue in Sec.~\ref{dynamic} by addressing how the lattice fluctuations alter the dynamic properties such as charge mobility, and the lifetimes of charge and lattice excitations. In each subsection, we specify the types of perovskites considered, ranging from zero-dimensional nanocrystals to their three-dimensional bulk counterparts. In Sec.~\ref{conclusion}, we conclude this with a summary and active areas of continued study.


\section{Static properties}
\label{static}

The generation, dissociation, and recombination of excitons determine the power conversion efficiencies and quantum yields of devices.  The energetics of excitons and other quasiparticle excitations offer figures of merit for designing semiconducting materials, as they correlate with relevant dynamical properties. Details of these static properties and how they depend on the static or evolving lattice are reviewed here. In particular we review (i) how lattice distortion influences exciton fine structure in nanocrystals and (ii) how the binding energies of excitons and their complexes are modulated by confinement and polaronic effects. 


\subsection{Exciton fine structure}

Lead halide perovskite nanocrystals are considered  a promising quantum light source material as they show a remarkable brightness for photoluminescence and high quantum yield. \cite{10.1021/acs.nanolett.9b00689} The origin of the bright emission has been a subject of debate, since there are conflicting experimental reports on the nature of the excitonic ground state, and the size of the dark-bright excitonic spittings\cite{10.1038/nature25147, 10.1021/acsnano.4c02905}. The excitonic fine structure of a material can be inferred in principle from spectroscopic measurements at cryogenic temperatures, though such measurements are challenging. While in bulk semiconductors the splitting of dark-bright exciton sublevels is very small, the splitting in nanocrystals can be substantial.

\begin {figure}
\centering\includegraphics [width=8.5cm] {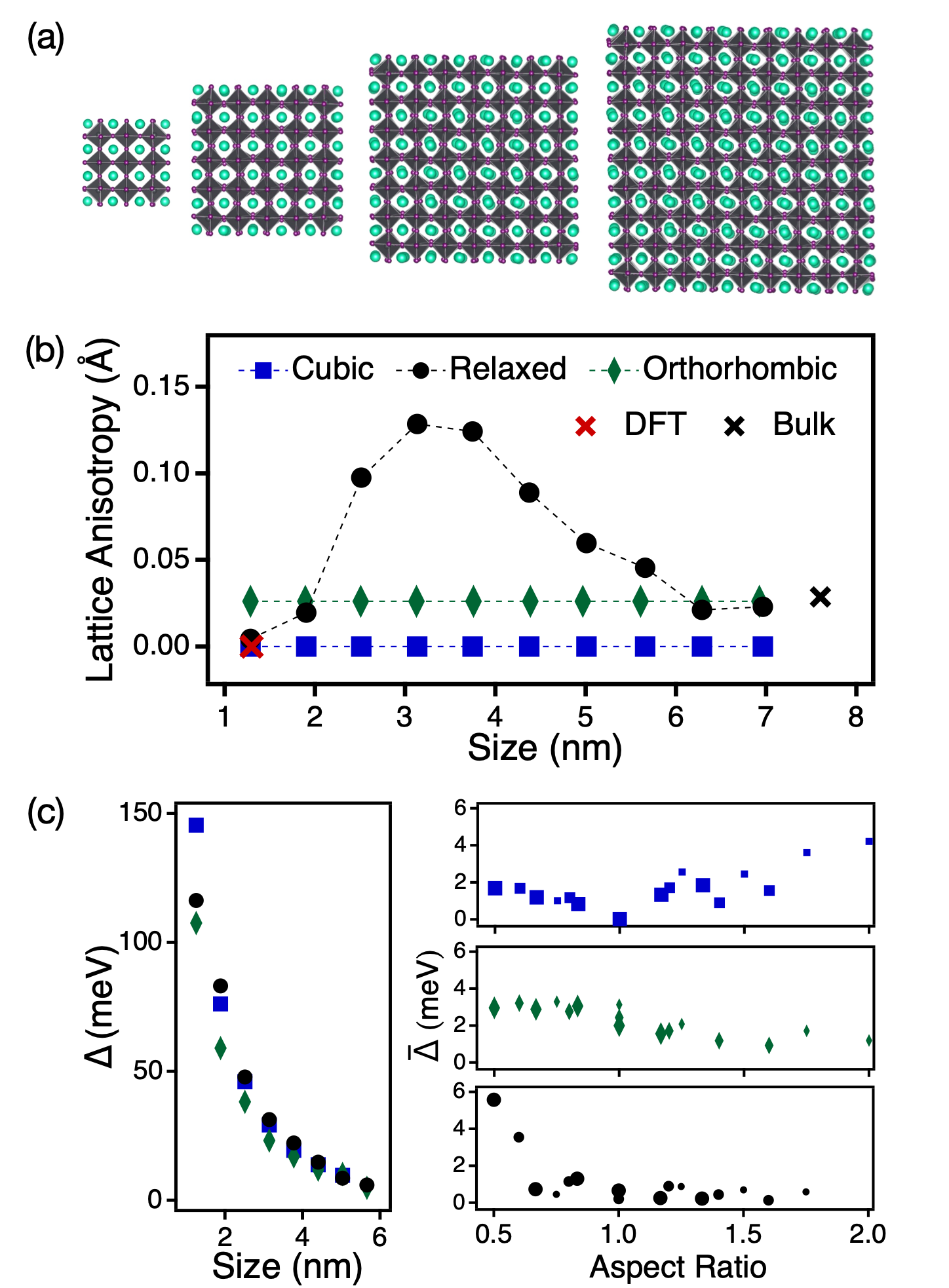}
\caption{(a) Relaxed structures of CsPbI$_3$ perovskite nanocrystals for the size of 1.9, 3.1, 4.4, and 5.7\,nm where I and Cs atoms are shown in purple and teal with gray Pb octahedra. (b) The lattice anisotropy parameter for cubic (blue squares), relaxed (black circles), and orthorhombic (green diamonds) structures. Red and black $\times$ symbols indicate the results from DFT and Bulk perovskite. (c) The splittings, $\Delta$, between dark and bright excitonic states with various nanocrystal sizes (left) and the standard deviation of splittings, $\bar{\Delta}$, between bright excitonic states as a function of aspect ratio (right) for cubic (blue), orthorhombic (green), and relaxed crystal structures (black). Reproduced with permission from ref \cite{10.1021/acs.nanolett.3c00861}. Copyright 2023 American Chemical Society.}
\label{Fig_NC}
\end{figure} 

For materials with strong spin-orbit coupling like the lead halide perovskites, spin and angular momentum are correlated, resulting in a spin lattice coupling that can change and possibly invert the spitting of optically bright and dark states. In the bulk, lead halide perovskites undergo structural phase transitions between low temperature orthorhombic phases, intermediate tetragonal phases, and high temperature cubic phases, each with different symmetry and thus changes to the electronic band structure. \cite{10.1021/acs.jpclett.7b00134} In nanoparticles, the locations of phase transformations are generally at different temperature than their bulk counterparts, and their significant surface-to-volume ratios can result in further complications as those surfaces relax and reconstruct.

Using an atomistic force field, \cite{10.1016/j.matt.2020.07.015} we studied a series of CsPbI$_3$ perovskite nanocrystals whose relaxed structures are shown in Fig.~\ref{Fig_NC}(a). We observed that the local lattice structure was size-dependent,  with the smallest nanocrystals stabilizing a highly symmetric cubic phase whereas the largest ones recovered a bulk-like orthorhombic structure. The degree of asymmetry was quantified in Fig.~\ref{Fig_NC}(b), using a lattice anisotropy parameter defined as the difference in  the Pb-Pb distance between shortest and longest axis of the nanocrystal. The values for cubic structures are zero and for orthorhombic structures, the tilting of the octahedra breaks the isotropic symmetry resulting in a small constant value of the order parameter. We found that the relaxed structures did not monotonically transition between cubic and orthorhomic with increasing size, and rather exhibit a maximum anisotropy for cubes with edge lengths of 2.7 nm. These observations were broadly consistent with experimental x-ray measurements.\cite{10.1021/acsenergylett.9b02395} This size-dependence stems from the competition between surface relaxation and the stability of the bulk orthorhombic phase. The establishment of the nanoparticle structure was only theoretically tractable with \emph{ab initio} derived forcefields, as it is computationally intractable to relax geometries directly with density functional theory for such large systems. 

The exciton fine structures were subsequently obtained from semiempirical pseudopotential methods \cite{10.1063/1.478431} using the Bethe-Salpeter equation \cite{neatonprl}  developed in Ref. \onlinecite{10.1021/acs.nanolett.3c00861} on the geometries obtained from the forcefield calculations. The electronic pseudopotentials were parameterized to include both the effect from lattice distortion and spin-orbit coupling. Only with such a theoretical framework could the correlated electronic structure of nanoparticles be computed. With this methodology, we found the excitonic ground state to be dark with spin triplet character, exhibiting a small Rashba coupling, which is implied by the similar splitting trends as increasing nanocrystal sizes for all phases shown in Fig.~\ref{Fig_NC}(c, left). For cubic phases of cubic nanoparticles, the three bright states are degenerate, but deviations locally in the unit cell or morphologically in the aspect ratio of the nanocrystal could break this degeneracy.
With the relaxed structure of each nanocrystal, Fig.~\ref{Fig_NC}(c, right panels) describes the standard deviation $\bar \Delta $ of splittings between bright excitonic states as a function of aspect ratio. Unlike the cubic and orthorhombic structures, the splittings with low aspect ratio are significant, unique feature of relaxed structures. This study confirmed that the lattice symmetry is vital in interpreting exciton fine structure of perovskite nanocrystals. 


\subsection{Exciton binding energy in bulk} 
\label{exbe}

While the absolute energies of excitons determine what states are populated thermally, the exciton binding energy marks the threshold to generate free charge carriers from an initial photoexcitation and determines the relative proportion of bound excitons to free charge carries. Bulk lead-halide perovskites have shown anomalously small values for exciton binding energies, typically reported as less than thermal energy at room temperature, attracting a great attention as a promising material for photovoltaic applications. \cite{herz2016charge} First principle estimates of exciton binding energies had failed to recover the anomalously small exciton \textcolor{black}{binding energies,\cite{10.1021/acs.jpclett.7b03286}} and while simple effective mass models could be made consistent with experiments, they typically required values of the dielectric constants that were difficult to \textcolor{black}{interpret.\cite{pbtheory}} The difficulty in establishing the exciton binding energy from such studies implicated the role of exciton-polarons and lattice deformation in the renormalization of observed binding energies. Such effects are not often accounted for in solid-state electronic structure. 

\begin {figure}[t]
\centering\includegraphics [width=8.5cm] {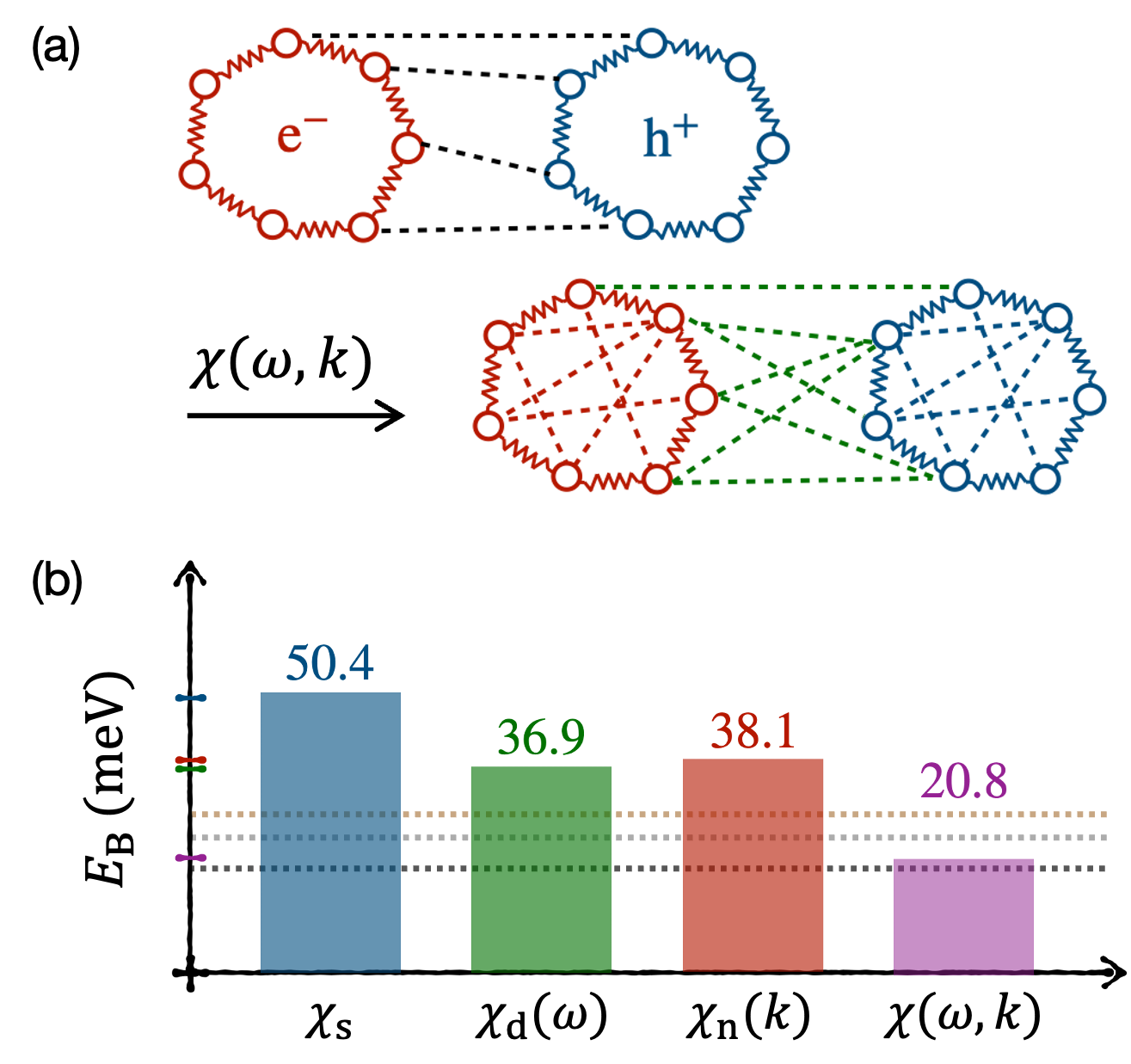}
\caption{(a) Illustration of two mechanisms describing how exciton binding energy is renormalized by the lattice. Reproduced with permission from ref \cite{parkjcp2022}. Copyright 2022 AIP Publishing LLC. (b) Exciton binding energy, $E_{\mathrm{B}}$, computed under different types of lattice screenings where dotted lines indicate experimental values from Ref. \cite{10.1039/c3ee43161d}(black), Ref. \cite{10.1038/ncomms6049}(gray), and Ref. \cite{10.1021/acs.jpclett.5b02099}(brown).  Reproduced with permission from ref \cite{park2022}. Copyright 2022 American Chemical Society.}  
\label{Fig_eb}
\end{figure}

The approach we developed was based on imaginary time path integrals\cite{feynman1, feynman2, chandler1981exploiting, ceperley1995path} using an explicit atomistic representation of the lattice. Perovskites have highly dispersive bands, so the low energy band structure is well described by effective mass models. The lattice Hamiltonian was given by the previously validated atomistic force field \cite{mattoni} and the charge-lattice coupling was described by the sum of pseudopotentials \cite{park2022, parkjcp2022}, in which way all orders of anharmonic features of perovskite lattice can be captured. Within this model, imaginary time path integrals could then be used to treat \textcolor{black}{electron-hole} correlation and charge-lattice coupling, without resorting to perturbation theory or approximate analytical results. 

We found good agreement with most experimental reports for exciton binding in the methylammonium lead-iodide perovskites, MAPbI$_3$, establishing a binding energy of 21 meV. The lattice renormalization of exciton binding occurs through two mechanisms, illustrated in Fig.~\ref{Fig_eb}(a). First, the lattice screens the direct interaction between charges by the static dielectric constant. This is a direct effect, weakening the Coulomb potential between the electron and hole. Second, the lattice stabilizes the free charges, altering their and the exciton's self energy. This effect reflects in polaron formation of free charges or exciton-polaron formation of the bound electron-hole pair. We have found that exciton-polaron binding energies are relatively small, as a charge neutral object couples weakly to the lattice, but the stabilization of both the electron and hole from polaron formation is significant, and directly responsible for the attenuation of the binding energy. 

We developed an effective field theory to decompose the contribution from the lattice on excitonic properties,\cite{park2022} through different approximations to the lattice Green's function, $\chi(k,\omega)$. In principle, $\chi(k,\omega)$ includes both a frequency, $\omega$ and wavevector, $k$, dependence. In a \textit{static} screening model, the electron-hole interaction is only screened by an effective dielectric constant, ignoring the dependence on both $\omega$ and $k$, $\chi(k,\omega)\approx \chi_\mathrm{s}$. In a \textit{dynamic} screening model, we could incorporate the effect from harmonic phonons through a Frohlich-like model \cite{frohlich2, frohlich1} for a charge that is coupled with polarization field produced from the collective harmonic motions of the lattice. The models in this approximation were dispersionless, so $\chi(k,\omega)\approx \chi_\mathrm{d}(\omega)$, but whose frequency dependence admits the formation of a polaron through the quantum mechanical treatment of the lattice. Alternatively, a \textit{nonlocal} screening model with spatially dependent dielectic function from the explicit MAPbI$_3$ lattice, $\chi(k,\omega)\approx \chi_\mathrm{n}(k)$, was developed. In this model, the frequency dependence was ignored, which is equivalent to assuming it is well-described classically.  The computed exciton binding energy under each type of lattice screening is summarized in Fig.~\ref{Fig_eb}(b), including the full lattice Green's function that incorporates both non-locality and frequency dependence. Comparisons show that the inclusion of lattice effects from either dynamic or nonlocal screening reduces the attraction between electron and hole by about 10 meV, but incorporating both of these effects provides an excellent agreement with experiments, \textcolor{black}{accounting for a nearly 30 meV red shift of the binding energy (21 meV) compared to the simple effective mass estimate (50 meV)}.\cite{galkowski2016determination} This clarified the origin of small exciton binding energy, emphasizing the importance of anharmonic lattice fluctuations and polaron formation in lead halide perovskites. 

\begin {figure}[t]
\centering\includegraphics [width=8.5cm] {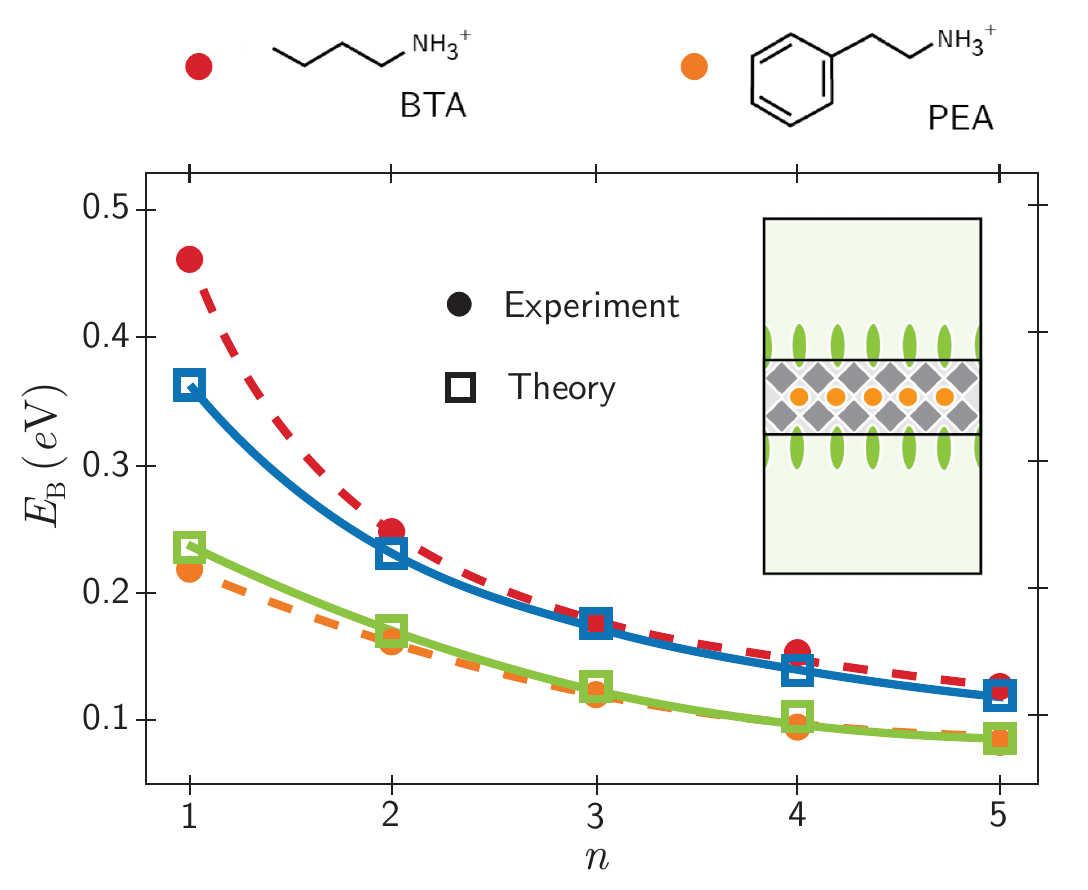}
\caption{Exciton binding energies for thin slab geometries, illustrated in the inset, for both (BTA)$_2$(MA)$_{n-1}$Pb$_n$I$_{3n+1}$ (red and blue) and  (PEA)$_2$(MA)$_{n-1}$Pb$_n$I$_{3n+1}$ (orange and green). Circles are experimental data from Ref. \cite{urban2020revealing} and squares are from theoretical calculations from  Ref. \cite{rana2025interplay}. Lines are guides to the eye.  Reproduced with permission from ref\cite{rana2025interplay}. Copyright 2025 American Chemical Society.} 
\label{Fig_ex2d}
\end{figure}
\subsection{Exciton binding in two-dimensions} 

Two-dimensional layered hybrid organic-inorganic perovskites have been proposed as alternatives to their bulk counterparts due to enhanced tunability and stability.\cite{leung2022stability} Consequently, the photophysics of two-dimensional perovskites are different than bulk perovskites. The heterogeneous layer composition leads to a dielectric contrast that enhances the electron-hole interaction.\cite{hanamura1988quantum} The electron-hole interaction is also enhanced by quantum confinement that arises from the exciton and inorganic layer length-scales being similar in value.\cite{hanamura1988quantum} However, it remained to be understood how excitons are renormalized due to charge-phonon interactions in two-dimensional perovskites. To approach this problem, we considered two model systems with organic ligands,  n-butylammonium (BTA) and phenylethylammonium (PEA), (BTA)$_2$(MA)$_{n-1}$Pb$_n$I$_{3n+1}$ and (PEA)$_2$(MA)$_{n-1}$Pb$_n$I$_{3n+1}$ where $n$ is the number of inorganic sublayers. A change in $n$ corresponds directly to a change in the quantum well thickness, thus also modifying the confinement. We studied both crystals formed from alternating slabs of organic and inorganic material, as well as single sheets dispersed in solution, the latter of which is report in Fig.~\ref{Fig_ex2d}.

To study the renormalization to exciton binding in these model systems and geometries, we developed a path integral framework\cite{rana2025interplay} similar to the one used to investigate bulk perovskites. Specifically, we employed continuum electrostatics along with the thin-film Frohlich interaction derived for a thin-film geometry.\cite{sio2022unified} Figure~\ref{Fig_ex2d} shows that the inclusion of a dynamic lattice brings exciton binding energies in reasonable agreement with experiment for both materials.\cite{urban2020revealing} The binding energy for the BTA ligand is much larger then the PEA ligand due to the large dielectic contrast between the organic layer and the inorganic lattice. Both exhibited a strong dependence on $n$, illustrating the effect of dielectric confinement.  When studying the polaron, we observed a non-monotonic behavior of polaron binding energy as a function of quantum well thickness $n$. We were able to capture this non-monotonic trend with a variational bound that has two length scales: one that corresponds to an effective dielectric screening radius and another that corresponds to the polaron radius.\cite{rana2025interplay} To physically rationalize this trend, we consider two competing mechanisms. The  
first is that as $n$ increases, the amount of polar inorganic material increases, thus leading us to expect the electron-phonon interaction to strengthen. The second is that as $n$ increases, a lattice dipole becomes weakened due to stronger screening from the inorganic layer, thus leading us to expect the electron-phonon interaction to weaken. Overall, we were able to elucidate the interplay between quantum confinement, dielectric confinement, and charge-phonon coupling in two-dimensional perovskites.


\subsection{Biexciton binding in cubic nanocrystals} 

While there is a reasonable agreement on the size of exciton binding energies in bulk perovskites, the values for perovskite nanocrystals are comparatively less established, and the understanding of multiparticle energies is even more unclear. In nanocrystals,  quantum confinement can lead to strong many-body interactions among photoexcited charges and these can significantly affect the optical properties, \cite{10.1002/adma.202208354} implying that it is necessary to study the behavior of exciton complexes such biexcitons. Recently,  experimental studies\cite{dana2021unusually} proposed an anti-binding behavior of biexcitons in CsPbBr$_3$ nanocrystals, implicating the potential lattice effect on these interactions. A number of biexciton binding energies for lead halide perovskites have been reported to date, whose values span a wide range even with opposite signs. \cite{dana2021unusually, 10.1002/adma.202208354} However, experimental analysis is difficult because of effects from thermal broadening, spectral drifts, and the large variance in the nanocrystal sizes. \cite{10.1021/acs.nanolett.1c01291} 

\begin {figure}
\centering\includegraphics [width=8.5cm] {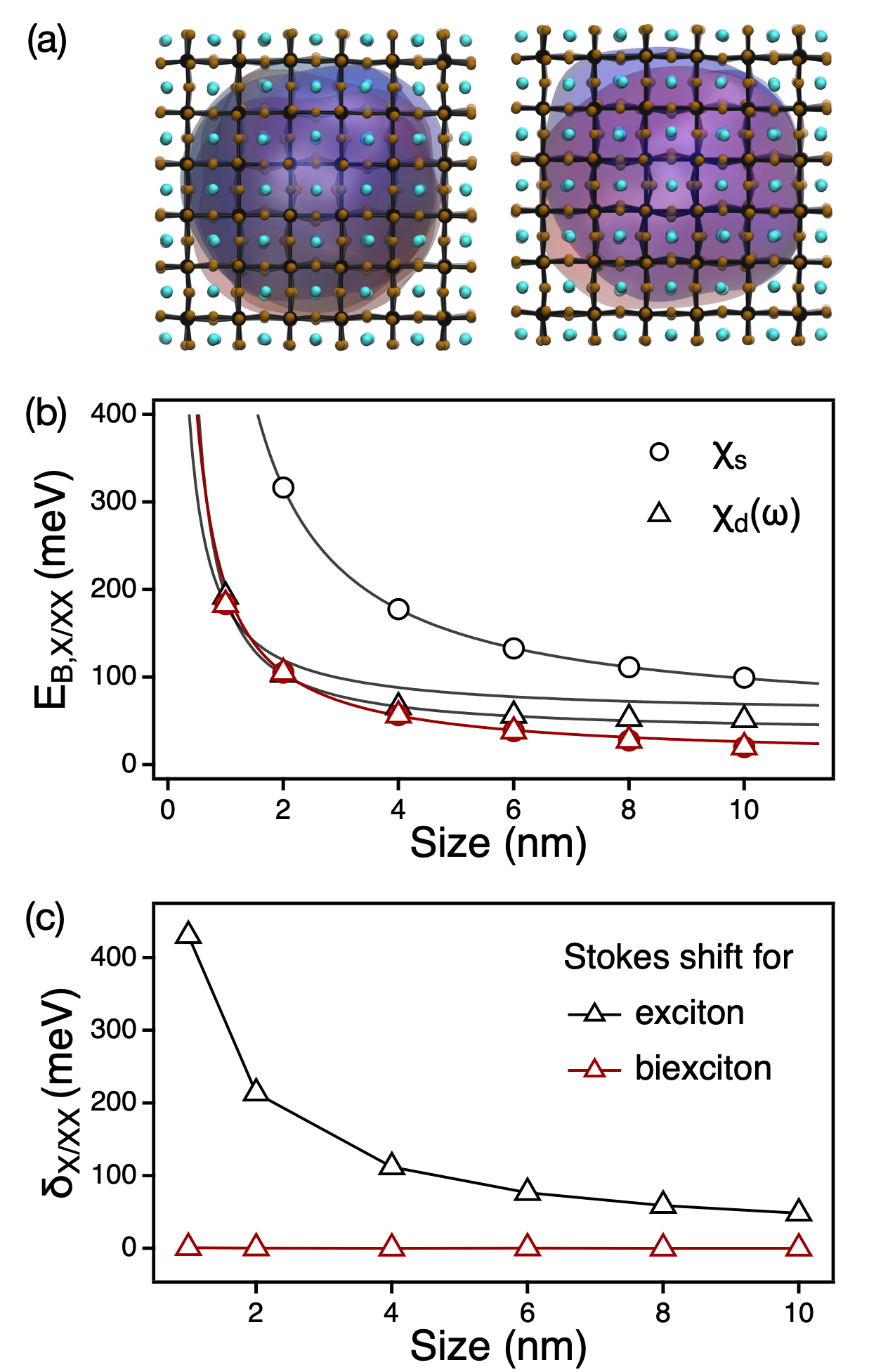}
\caption{(a) Simulation snapshot of biexciton (left) and exciton (right) interacting with CsPbBr$_3$ perovskite nanocrystal with 3.56 nm as the edge length of nanocrystal. Electrons and holes are represented as red and blue ring polymers. (b) Exciton ($E_{\mathrm{B,X}}$, black) and biexciton ($E_{\mathrm{B,XX}}$, red) binding energies, over the range of nanocrystal sizes where results from static and dynamic lattice screenings are shown with circles and triangles, respectively. Solid lines are fitted lines. (c) The difference in exciton ($\delta_{\mathrm{X}}$, black) and biexciton ($\delta_{\mathrm{XX}}$, red) binding energies from static screening, implying the values of Stokes shift. Solid line is a guide to the eye.
 Reproduced with permission from ref \cite{park2023}. Copyright 2023 American Physical Society.} 
\label{Fig_biex}
\end{figure}

Adopting the same path integral approach as employed to study other perovskite systems, the energetics of quasiparticle excitations in CsPbBr$_3$ perovskite nanocrystals are reviewed \cite{park2023} where Fig.~\ref{Fig_biex}(a) shows the snapshots of molecular dynamics simulations of biexciton and exciton in CsPbBr$_3$ nanocrystal with electrons and holes represented as imaginary time paths \textcolor{black}{under the external potential accounting for a dielectric confinement effects}. For comparison, the static and dynamic lattice screening models noted above are also considered. Over the range of nanocrystal sizes, exciton and biexciton binding energy calculations have been performed under each type of lattice screening.
Shown in \textcolor{black}{Fig.~\ref{Fig_biex}(c)} is the difference in exciton and biexciton binding energies compared with results from static screening model. Even with the strong size dependence of both exciton and biexciton binding energies, the large values of $\delta_{\mathrm{X}}$ imply that the inclusion of lattice effects either from dynamic or explicit CsPbBr$_3$ lattice screening largely suppress the exciton binding energy, whereas lattice effects are negligible for biexciton binding. This observation is rationalized by the mechanism of polaron formation where instead of weakening the electron-hole interaction, lowering the self-energies of free charges contributes dominantly to reducing the binding energies in nanocrystals, which is not relevant for biexciton binding energies. 
Further, the calculations on biexciton binding energy, in agreement with other size-dependent measurements on CsPbBr$_3$ nanocubes, \cite{10.1021/acs.jpclett.0c01407} confirm that biexcitons are bound for all nanocrystal sizes considered, regardless of the type of lattice screenings. 


\section{Dynamic properties}
\label{dynamic}

The dynamics of charges and the surrounding lattice play a crucial role in the functioning of semiconductor based devices. After photoexcitation, the subsequent relaxation dynamics involves a number of different processes which occur on timescales that span the picoseconds of exciton dissociation and lattice vibration, to nano- and microseconds for charge carrier lifetimes all of which conspire to determine a device's efficiency. Establishing a microscopic understanding of these processes, and how each are influenced by the soft, polar lattice, provides means of designing materials to optimize optoelectronic properties. In this section, we focus on how lattice fluctuations alter the diffusion and lifetimes of charge carriers as well as the lifetime of lattice vibrations. 


\subsection{Charge carrier mobility}

In the application of semiconducting materials, mobility of charge carriers is one of primary properties which limits the performance of materials. In photovoltaic devices, free charge carriers need to reach the extraction layer within their lifetime to make use of them whereas in light-emitting devices, larger mobility implies the greater brightness with low driving voltage. \cite{10.1038/nphoton.2016.62} Experimentally, the relative carrier mobility can be extracted from the photoinduced conductivity measurement using the pump-probe THz transmission dynamics after photoexcitation where the intensity of the signal is proportional to the density of free charge carriers. \cite{adma.201305172} Studying charge transport phenomena in lead halide perovskites, Wehrenfennig \textit{et al.} showed that the difference in charge mobility highly affects the performance of materials. \cite{adma.201305172} 
Charge lattice interactions and polaronic effects in particular have been implicated in observations of renormalizing charge mobilities, as scattering off of the lattice is the primary mechanism of dissipation for excess charges\cite{mayers2018lattice}. 

\begin {figure}[t]
\centering\includegraphics [width=8.0cm] {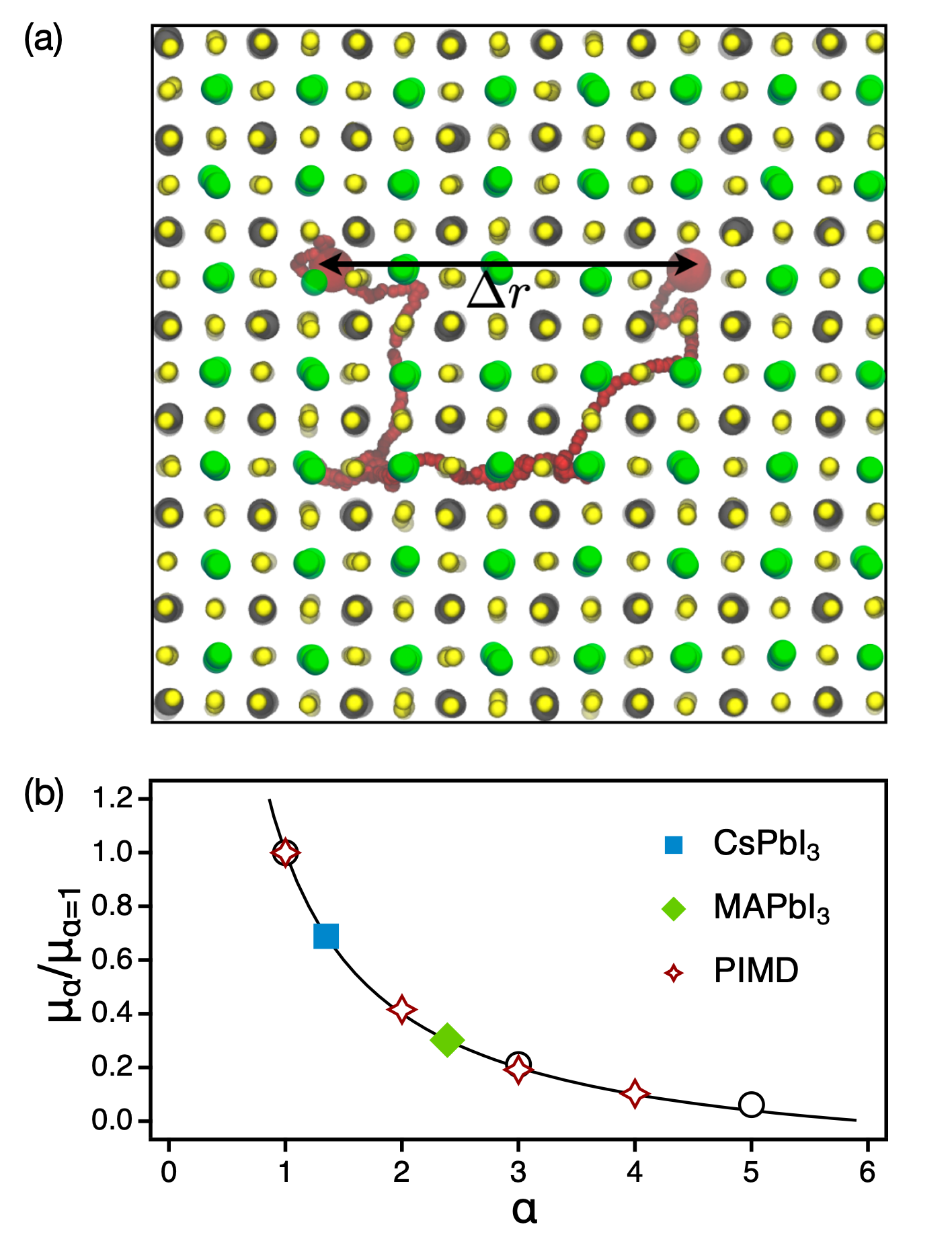}
\caption{\textcolor{black}{(a) Schematic of how to evaluate effective mass from an open-chain imaginary time path.
(b) The relative charge mobilities $\mu$ obtained from effective masses at various electron-coupling strength $\alpha$ in CsPbI$_3$ and MAPbI$_3$ perovskites with results from path integral approach. Solid line represents the estimate from Feynman's variational method.}
}
\label{Fig_mobility}
\end{figure}

Using our quasiparticle path integral approach, \cite{feynman1, feynman2, chandler1981exploiting, ceperley1995path} we have estimated the relative charge carrier mobility at various electron-phonon coupling strengths using Frohlich model Hamiltonian. \cite{frohlich2, frohlich1} In the anti-adiabatic limit which is applicable to lead halide perovskites, the charge mobility $\mu $ can be obtained from their renormalized mass $m^*$ as \cite{10.1103/physrevb.107.115203}
\begin{equation}
\mu = \frac{e}{2 \alpha \omega m^*} e^{\beta \hbar \omega}
\end{equation} 
where $\omega$ is a longitudinal optical frequency of phonons, $\alpha$ is a dimensionless Frohlich coupling constant, $\beta$ is inverse temperature times Boltzmann's constant and $\hbar$ is Planck's constant. An inverse renormalized mass at a given electron-phonon coupling strength can be computed using a linear response relation \cite{emass1}
\begin{equation}
\frac{1}{m^*} = \lim_{\beta \rightarrow \infty} \frac{\langle (\Delta r )^2\rangle }{3 \beta \hbar^2}
\end{equation} 
where $\Delta r$ is the distance between the first and the last beads of an open-chain imaginary time path, pictorially shown in Fig.~\ref{Fig_mobility}(a). 

Adopting the material properties of bulk lead-halide perovskites, the relative charge mobilities are described in Fig.~\ref{Fig_mobility}(b), at various electron-phonon coupling strengths. In this parameter regime, the effective mass is well described using first order perturbation theory \cite{ptemass} and is consistent with Feynman's  variational method \cite{Fvarm}. Combining these two results, we were able to infer that the coupling with phonons makes the charge carriers heavier, while suppressing the carrier mobility, with a greater reduction expected for MAPbI$_3$ compared to CsPbI$_3$ due to its smaller Frohlich coupling constant. Molecularly, the weaker charge-lattice coupling for the inorganic lattice originates in the slightly less polarizable A-site cation.

\subsection{Electron-hole recombination rate}

At typical operating conditions for bulk photovoltaics, electron-hole recombination determines the lifetimes of charge carriers. Despite their modest mobilities, the lead halide perovskites exhibit large diffusion lengths, enabling exception power conversion efficiencies. Over the last decade, it has been found that the origin of the large diffusion lengths is an abnormally long free carrier lifetime, or correspondingly low radiative recombination rate.\cite{science.aaa5333} 
It had been conjectured that polaronic effects play a major role in the underlying mechanism of low electron-hole recombination rates of perovskites, \cite{Zhu:2015eb4, ncomms12253} though direct evidence theoretically or experimentally had been lacking. We were able to elucidate the molecular origin of the long charge carrier lifetime as originating in an emergent repulsion between electrons and holes mediated by the soft perovskite lattice. 

In general, the charge recombination rate in the bulk is understood using a bimolecular rate expression. \cite{adma.201305172} For the reaction of $e^- + \mathrm{h}^+ \rightarrow \hbar \nu $, the bimolecular recombination rate $k_{\mathrm{r}}$  can be defined as 
\begin{equation}
\frac{d n_{\mathrm{e}}}{d t} = - k_{\mathrm{r}} n_{\mathrm{e}} n_{\mathrm{h}}
\end{equation}
where $n_{\mathrm{e/h}}$ is the electron/hole density. We used quasiparticle path integral molecular dynamics simulations  with an atomistic description of MAPbI$_3$ lattice to evaluate this rate. In a bulk system, recombination is a bimolecular process, so the rate is proportional to the probability of \textcolor{black}{bound electron and hole being co-localized}, and a conditional probability of radiating provided they are \textcolor{black}{co-localized. \cite{parkjcp2022, park2022}} Both depend on the effective potential between charges, as mediated through a finite temperature, fluctuating lattice. Using explicit simulations, we found at 300 K a non-local lattice Green's function,  $\chi_\mathrm{n}(k)$ was well approximated by
\begin{equation}
\chi(k) =\frac{\chi_0}{1 - l_s^2\mathbf{k}^2 + l_s^2 l_c^2 \mathbf{k}^4} 
\label{chik}
\end{equation}
characterized by two lengthscales, $l_s$, and $l_c$ which are screening and correlation lengths associated with dielectric fluctuations. This functional form includes a single resonant peak that results from the double well potential of the optical mode.\cite{multiphonon} 

\begin {figure}
\centering\includegraphics [width=8.5cm] {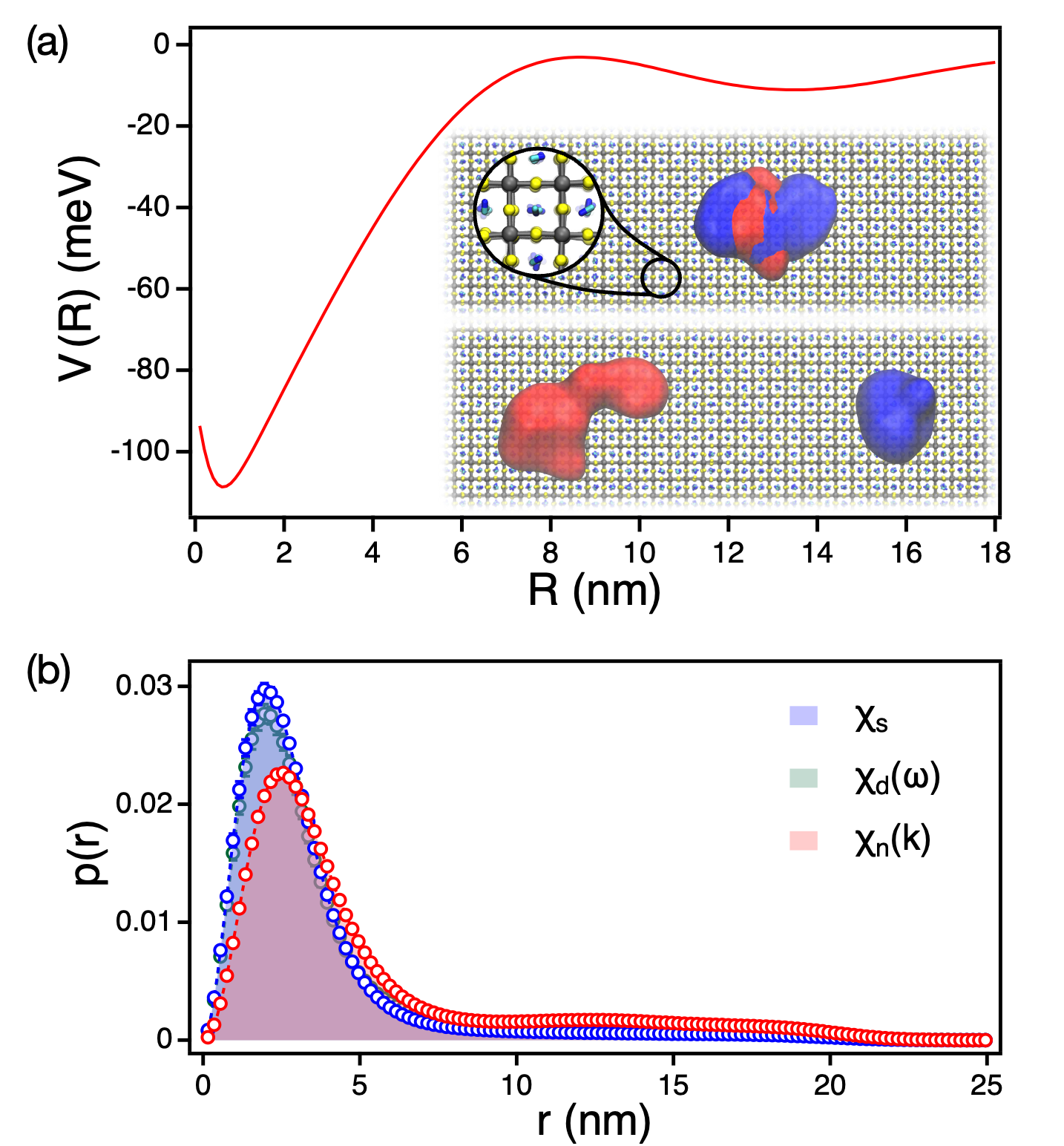}
\caption{(a) An effective potential between electron and hole in MAPbI$_3$ perovskite as a function of the distance between centroids of quasiparticles. The inset is the representative snapshots of molecular dynamics simulations of electron (red) and hole (blue) with the MAPbI$_3$ lattice where Pb$^{2+}$, I$^{-}$, and MA$^+$ are shown in gray, yellow, and blue atoms, respectively. (b) The electron and hole radial probability distribution under different types of lattice screenings from path integral simulations. Reproduced with permission from ref \cite{park2022}. Copyright 2022 American Chemical Society.} 
\label{Fig_rate}
\end{figure}

As the lattice Green's function screens in the electron-hole interaction, in the limit that the lattice is well described classically, it implies an effective potential 
\begin{equation}
\begin{aligned}
V(r) = -\frac{e^2}{4\pi \varepsilon_0 r} \left [ 
\frac{1}{\varepsilon_{\infty}} + \frac{1}{\varepsilon^*} + \frac{\gamma}{4\delta \varepsilon^*}
e^{-r\delta}\sin[r\gamma - \theta ] \right ]
\end{aligned}
\label{Veff}
\end{equation}
where \textcolor{black}{$r$} is the distance between two charges, \textcolor{black}{$1/\varepsilon^* = \chi_0 \alpha^2 \varepsilon_0/e^2$}, $\gamma^{-1}\approx \sqrt{2}\ell_c$, $\delta^{-1} \approx 2\ell_c/\sqrt{1-\ell_s/2 \ell_c}$,  and $\theta = \arctan[2\delta / \gamma]$. Notably this potential, shown in \textcolor{black}{Fig.~\ref{Fig_rate}(a)}, includes a barrier at intermediate distances. This electron-hole repulsion significantly decreases their wavefunction overlap relative to expectations without a fluctuating lattice, $\chi_\mathrm{s}$, or within its dispersionaless harmonic approximation, $\chi_\mathrm{d}$. All three approximations to the screening from the lattice are shown in  \textcolor{black}{Fig.~\ref{Fig_rate}(b)} where the electron-hole radial probability density is plotted, $p(r)$. Under standard approximations using either the static or harmonic approximation, the expected lifetimes of charge carriers at concentrations $n_{\mathrm{e}} = 10^{17} / \mathrm{cm}^3$ are on the order of 1/$k_{\mathrm{r}} n_{\mathrm{e}} \approx$10 ns, while it rises to 1/$k_{\mathrm{r}} n_{\mathrm{e}} \approx$200 ns when the nonlocal screening from the lattice is taken into account. This latter estimate was in very good agreement with photoluminescence lifetime measurements under similar \textcolor{black}{carrier concentrations.\cite{science.aaa5333, herz2016charge}} This study thus resolved the longstanding question concerning the anomalously long carrier lifetimes, by invoking a novel lattice mediated repulsion. This mechanism was molecular in origin, as reflective in the lengthscales $l_s$, and $l_c$.

\subsection{Radiative recombination in nanocrystals}

\begin {figure}
\centering\includegraphics [width=8.5cm] {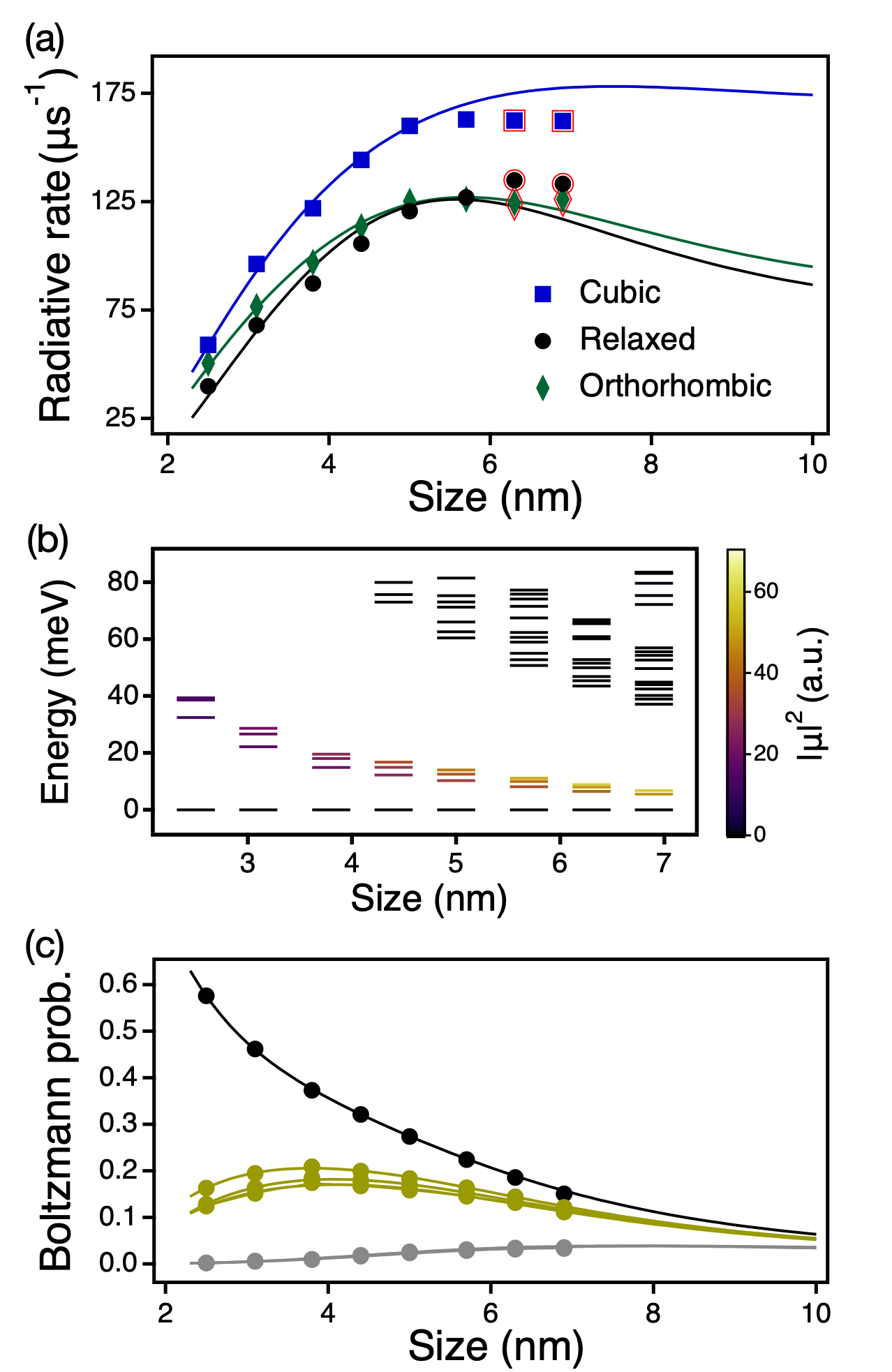}
\caption{(a) Size-dependent radiative recombination rates show non-monotonicity due to the increasing oscillator strength for larger nanocrystals in competition with increasing thermal population of higher energy, dim excitonic states. Points represent rates calculated on atomistic nanocrystal structures, and the lines represent radiative rates extrapolated from fits of exciton energies and oscillator strengths; no rates were fitted. The red-ringed points are test points and were not included in the fitting data. 
(b) Representative exciton energy level diagram for orthorhombic CsPbI$_3$ nanocrystals of different edge lengths. The color corresponds to the squared transition dipole magnitude of the exciton. 
(c) Size-dependent Boltzmann probabilities for the lowest exciton (black), bright triplets (gold), and dim states (grey) of orthorhombic CsPbI$_3$ show nonmonotonicity due to the thermal population of higher-lying states at larger sizes.
Reproduced with permission from ref \cite{abbas2024non}. Copyright 2025 The Authors.
}  
\label{Fig_rateNC} 
\end{figure}

The optoelectronic properties of nanocrystals depend intricately on both size-dependent quantum confinement and lattice symmetry. When a photogenerated electron-hole pair forms, its spatial extent - characterized by the exciton Bohr radius - determines how confinement affects the excitonic behavior. In nanocrystals of radius smaller than the Bohr radius, the exciton is confined within the nanocrystal volume, leading to discrete energy levels and size-tunable radiative recombination. These recombination dynamics are further modulated by the underlying lattice symmetry.\cite{alivisatos_semiclusters1996, protesescu2017} Room temperature photoluminescence measurements of perovskite nanocrystals reveal fast radiative recombination on a nanosecond timescale with near-unity photoluminescence quantum yield, indicating possible applications in quantum light generation.\cite{protesescu2017} Recent studies of size-dependent radiative rates in confined nanocrystals have revealed a non-monotonic trend in exciton radiative recombination for CsPbBr$_3$ nanocrystals, with peak rates observed for nanocrystals around 5.5 nm in size — corresponding to an intermediate confinement regime. This nonmonotonic behavior is most pronounced in structures distorted away from cubic symmetry due to modification of the exciton level structure under reduced lattice symmetry \cite{abbas2024non} Thus, a detailed understanding of how quantum confinement and lattice distortion influence radiative recombination is necessary for designing optimal platforms for perovskite nanocrystal-based optical devices. \\

To elucidate the interplay of structural and electronic effects, we combined molecular dynamics simulations for lattice dynamics with atomistic semi-empirical pseudopotential calculations for excitonic states. The detailed theoretical framework has been described elsewhere.\cite{10.1016/j.matt.2020.07.015, 10.1021/acs.nanolett.3c00861}  A size-dependent transition was found in agreement with previous work,\cite{10.1021/acs.nanolett.3c00861} where small, approximately 2 nm CsPbX$_3$ cubes are cubic and larger, $>$ 4 nm CsPbX$_3$ nanocrystals (X = I, Br) have increased lattice distortion toward the orthorhombic geometry. 
Atomistic pseudopotential + BSE calculations were carried out for lead halide perovskite nanocrystals ranging from 2 nm to 7 nm in size, considering three structural symmetries: ideal cubic, orthorhombic, and force-field-relaxed geometries. The computed energies and oscillator strengths were fitted to extrapolate the radiative rate at larger nanocrystal sizes. Boltzmann thermal averaged radiative rates were computed in the time-dependent perturbation theory framework as
\begin{equation}
    \langle k_r \rangle = \left[\sum_{n} e^{- \beta \hbar \omega_n} \frac{\omega_n^3 |\mu_n|^2}{3 \pi \epsilon_0 \hbar c^3} \right]/ \sum_{n} e^{- \beta \hbar \omega_n}
    \label{Eq_avgRate}
\end{equation}
where $\mu_n$ and $\omega_n$ are the oscillator strength and Bohr frequency of exciton $n$, $\beta = k_B T$, $\epsilon_0$ is the permittivity of free space, and $c$ is the speed of light. Radiative rates are proportional to the exciton oscillator strength, which increases with nanocrystal size due to the larger electron-hole overlap in the nanocrystal volume. From this trend, it is expected that the radiative rates increase monotonically for larger nanocrystals. However, as shown in Fig. \ref{Fig_rateNC}(a), the radiative rates increase between 2–5 nm, which we attribute to a size-dependent increase in the oscillator strength of the lowest bright exciton state, Fig \ref{Fig_rateNC}(b). This trend is observed irrespective of crystal symmetry. However, this increase in oscillator strength does not continue to reduce the radiative lifetime in larger nanocrystals approaching the weak confinement regime. Instead, as nanocrystal size increases, the growing density of states enables thermal population of higher-lying, low oscillator strength "dim" exciton states. This thermal redistribution reduces the Boltzmann population of the lowest bright triplet manifold Fig \ref{Fig_rateNC}(c), leading to longer radiative lifetimes despite stronger individual transitions. These dim states exhibit an equilibrium population distribution due to the long, nanosecond timescale of radiative recombination compared to the picosecond hot exciton cooling rates. This thermal redistribution reduces the Boltzmann population of the lowest bright triplet manifold, leading to longer radiative lifetimes despite stronger individual transitions. The observed nonmonotonic radiative rates emerge distinctly in orthorhombic and relaxed nanocrystals of lowered structural symmetry due to exciton level splitting that simultaneously reduces the energy gap from the bright to dim states and increases the density of the lower energy dim states; cubic nanocrystals are not predicted to show non-monotonic radiative rates.

\subsection{Phonon dephasing}

\begin {figure}
\centering\includegraphics [width=8.6cm] {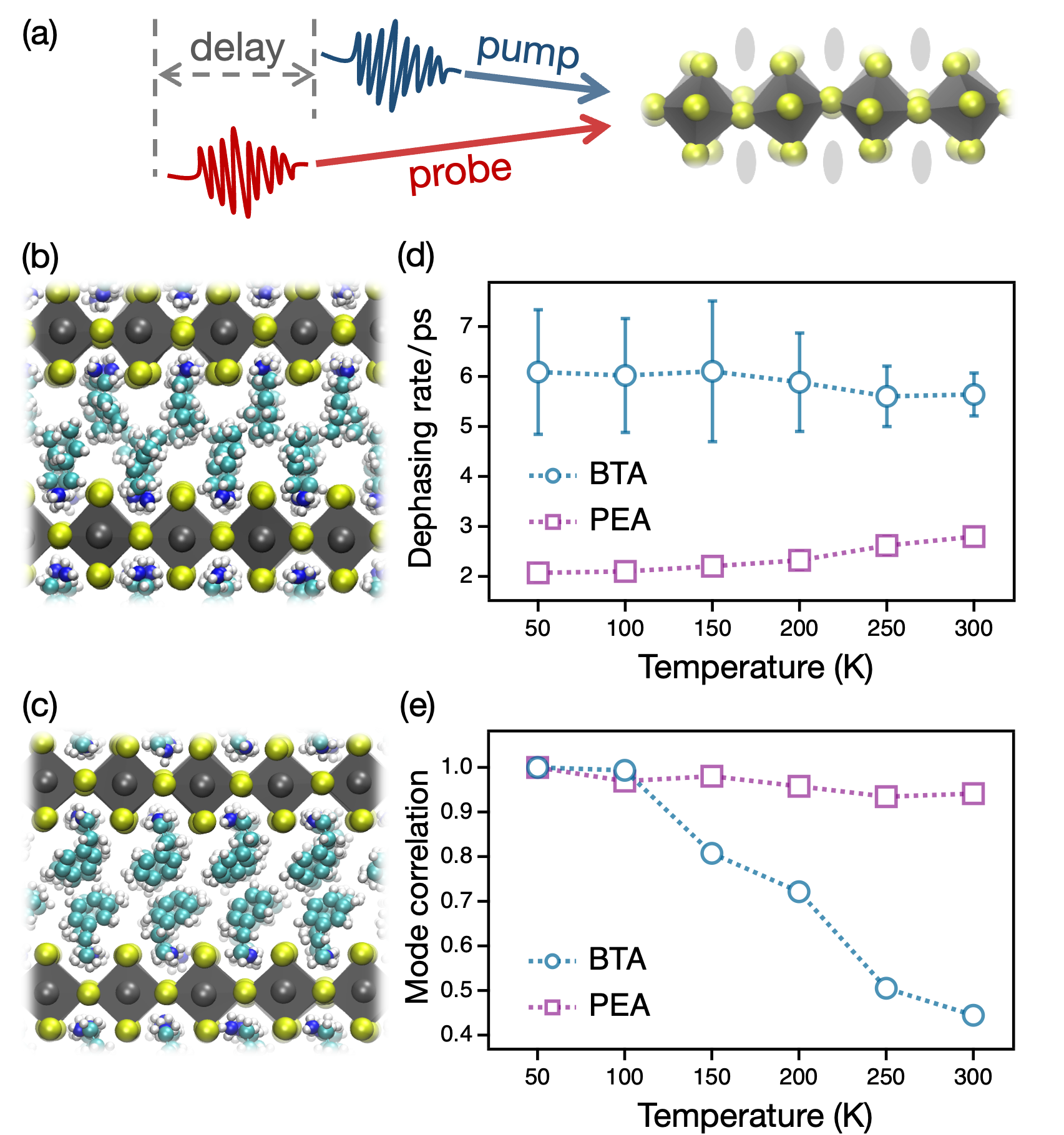}
\caption{(a) Schematic illustration of transient pump-probe measurement (absorption spectroscopy) on two-dimensional layered perovskites. (b and c) Snapshots of molecular dynamics simulations on (b) (BTA)$_2$PbBr$_4$ and (c) (PEA)$_2$PbBr$_4$ layered perovskites, respectively. (d) Phonon dephasing rate of the lowest optical mode with BTA (blue circles) and PEA (purple squares) organic ligands. (e) Correlation between lowest frequency optical modes at different temperatures. 
Dotted lines are guide to the eye. Reproduced with permission from ref \cite{park2021}. Copyright 2021 National Academy of Sciences.} 
\label{Fig_dephasing}
\end{figure}

The relaxation of excited charges proceeds through charge-phonon interactions, exciting lattice vibrations which themselves must subsequently relax. The examination of the dynamics of the lattice vibrations that charges coupled to can thus provide insight into the behavior of excited states and clarify the ways in which they are coupled. In layered perovskites, experimental observations have reported a rich lattice dynamics owing to their ionic inorganic framework together with the softness arising from the organic barrier layers. Early work by Quan \textit{et al.} reported the dependence of photoluminescence quantum yield on the organic ligand environments. \cite{park2021} They observed that layered perovskites with aromatic organic groups show much higher yields compared to those with alkyl groups. Delving into the role of dynamics of the lattice, with a pump-probe spectroscopy (illustrated in Fig.~\ref{Fig_dephasing}(a)), vibrational relaxation dynamics were studied through phonon dephasing rate with mainly two different organic ligands, n-butylamine (BTA)  and phenylethylamine (PEA). 
It was found that phonon relaxation with BTA ligands is much faster than with PEA ligands. While with the faster dephasing rate over the range of temperature, contrary to the prediction from a general scattering theory, \cite{PRBverma} the dephasing rates of BTA ligands exhibit no temperature dependence whereas the rates from PEA ligands show a linear temperature dependence. Hence, these distinct dynamics depending on the organic ligands given the same inorganic framework indicate the importance of the organic ligands in describing lattice dynamics in layered perovskites.  

To obtain a molecular level understanding of vibrational dynamics of layered perovskites, molecular dynamics simulations \cite{ffref} were performed. While simple, the basic physical properties such as lattice constants and mechanical properties were reasonably reproduced by the model whose simulation snapshots with BTA and PEA ligands are shown in Fig.~\ref{Fig_dephasing}(b) and (c), respectively. To effectively extract the phonon modes, the fluctuation-dissipation theorem is employed to relate displacement correlations from simulations to the dynamical matrix \cite{LKong1} where the analysis on the resultant vibrational frequencies and phonon modes verifies that the organic ligands largely participate in forming vibrational modes in layered perovskites. \cite{park2021} 

For the lowest frequency optical mode, the atomistic model qualitatively captured the phonon dephasing dynamics observed experimentally in each layered perovskites. In the classical limit, derived from the Fermi's golden rule, the relaxation rate of vibrational mode $\lambda$ can be computed from the Fourier transform of the corresponding force-force correlation function \cite{egorov1999quantum} as 
\begin{equation}
\Gamma_{\lambda} = \frac{1}{k_{\mathrm{B}}T} \int_0^{\infty} dt \cos (\omega_{\lambda} t) \langle F_{\lambda}(0) F_{\lambda}(t) \rangle
\end{equation}
where $\omega_{\lambda}$ is the optical frequency and $F_{\lambda}$ is the force exerted on mode $\lambda$ due to the coupling with their surrounding lattice. Figure~\ref{Fig_dephasing}(d) describes the dephasing rate in BTA and PEA layered perovskites over a range of temperatures, showing that the dephasing rates of BTA layered perovskite is higher than the one from PEA layered perovskite and are insensitive to the temperature. 

Different phonon dynamics in layered perovskites were understood by comparing their packing geometry and anharmonicity of organic ligands. As deduced from Fig.~\ref{Fig_dephasing}(b) and (c), compared to PEA ligands where molecules are well aligned via strong $\pi-\pi$ interaction, BTA ligands hold amorphous packing geometry even at low temperature, whose dynamic disorder can rationalize the temperature insensitive relaxation rate. In terms of higher rate of BTA perovskites, this can be resolved by the large anharmonicity of BTA ligands. Based on the structural analysis of organic ligands, the local potential associated with the ligand orientation turns out to be highly anharmonic in BTA perovskite, showing a large deviation from the prediction based on fluctuation-dissipation theorem (see Ref. \cite{park2021} for the detailed analysis). This is further supported by calculations showing the correlation between optical modes at different temperatures, described in Fig.~\ref{Fig_dephasing}(e) where the value ranges from 0 to 1 with the value 1 indicating that two vibrational modes are identical. While for PEA perovskites, the values  are close to 1 across the whole range of temperature, the large drop in BTA perovskites as increasing the temperature implies that the optical modes are highly mixed with increasing temperature, a signal of large anharmonicity. Given the huge contribution of the organic ligands to the optical modes, the stronger anharmonicity from BTA perovskites justify the higher rate, facilitating the vibrational relaxation. 

%
%

\section{Concluding remarks}
\label{conclusion}

In this Account, we have outlined our studies describing the effects from the perovskite lattice on the excited state properties. For both static and dynamic properties, while various computational and analytical descriptions for excited states and lattice are considered in our approaches presented here, our focus has been on how to include the lattice effect and how important it is. Our calculations confirm that the lattice distortion resulting from the structural relaxation needs to be considered to properly describe the electronic states in perovskites. Further, it is found that the lattice fluctuations affect phonon dynamics itself and also dynamics of charge carriers, generically reducing the mobility, exciton binding energy, and electron-hole recombination rate, whereas there is negligible effects on biexciton bindings. We expect these results along with the strategic approaches drive the continued exploration on the role of lattice effects in elucidating the origin of properties in lead halide perovskites. 

\vspace{1mm}
\section{ACKNOWLEDGMENTS}
Support for this work was provided by the U.S. Department of Energy, Office of Science, Office of Basic Energy Sciences, Materials Sciences and Engineering Division, under Contract No. DEAC02-05-CH11231 within the Fundamentals of Semiconductor Nanowire Program (KCPY23). Computational resources were provided in part by the National Energy Research Scientific Computing Center (NERSC), a U.S. Department of Energy Office of Science User Facility operated under contract no. DEAC02-05CH11231. We would like to thank other collaborators who participated in the work reviewed here including Amael Obliger, Daniel Weinberg, Peidong Yang, and Paul Alivisatos. 

\bibliographystyle{unsrt}
\bibliography{main}
\end{document}